\documentclass[showpacs,preprintnumbers,twocolumn,10pt,prl]{revtex4}
\usepackage{amsmath}
\usepackage{dcolumn}
\usepackage{bm}
\usepackage{graphicx}
\usepackage{amsfonts}
\usepackage{amssymb}
\usepackage{bbm}
\usepackage{float}
\usepackage[dvipdfm]{hyperref}

\begin{document}

\title{Quantum-memory-assisted entropic uncertainty principle under noise}
\author{Z. Y. Xu $^{1}$}
\email{zhenyuxu@suda.edu.cn}
\author{W. L. Yang $^{2}$}
\author{M. Feng $^{2}$}
\email{mangfeng@wipm.ac.cn}
\affiliation{$^{1}$School of Physical Science and Technology, Soochow University, Suzhou
215006, China \\
$^{2}$State Key Laboratory of Magnetic Resonance and Atomic and Molecular
Physics, Wuhan Institute of Physics and Mathematics, Chinese Academy of
Sciences, and Wuhan National Laboratory for Optoelectronics, Wuhan 430071,
China}

\begin{abstract}
The measurement outcomes of two incompatible observables on a particle can
be precisely predicted when it is maximally entangled with a quantum memory,
as quantified recently [Nature Phys. \textbf{6}, 659 (2010)]. We explore the
behavior of the uncertainty relation under the influence of local unital and
nonunital noisy channels. While the unital noises only increase the amount
of uncertainty, the amplitude-damping nonunital noises may amazingly reduce
the amount of uncertainty in the long-time limit. This counterintuitive
phenomenon could be justified by different competitive mechanisms between
quantum correlations and the minimal missing information after local
measurement.
\end{abstract}

\pacs{03.65.Ta, 05.40.Ca, 03.65.Yz}
\maketitle

One of the most remarkable features of quantum mechanics is the restriction
of our ability to simultaneously predict the measurement outcomes of two
incompatible observables with certainty, which is called Heisenberg's
uncertainty principle \cite{UP}. Nowadays, the more modern approach to
characterize the uncertainty principle is the use of entropic measures
rather than with standard deviations \cite{entropyU}. If we denote the
probability of the outcome $x$ by $p(x)$ when a given quantum state $\rho $
is measured by an observable $X$, the Shannon entropy $H\left( X\right)
=-\sum_{x}p(x)\log _{2}p(x)$ characterizes the amount of uncertainty about $%
X $ before we learn its measurement outcomes \cite{book QIC}. For two
non-commuting observables $Q$ and $R$, the entropic uncertainty relation can
be expressed as $H\left( Q\right) +H\left( R\right) \geq \log _{2}\frac{1}{c}
$ \cite{entropyU}, where $c=$ max$_{\alpha ,\beta }\left\vert \left\langle
\phi _{\alpha }\right\vert \varphi _{\beta }\rangle \right\vert ^{2}$ with $%
|\phi _{\alpha }\rangle $ and $|\varphi _{\beta }\rangle $ the eigenstates
of $Q$ and $R$, respectively. Since $c$ is independent of the states of
system to be measured, the widely studied entropic uncertainty relation
provides us with a more general framework of quantifying uncertainty than
the standard deviations (See a review in \cite{EUsurvey}).

However, the entropic uncertainty relation may be violated if a particle is
initially entangled with another one \cite{Violated}. In the extreme case,
an observer holding the particle $A,$ maximally entangled with particle $B$
(quantum memory), is able to precisely predict the outcomes of two
incompatible observables $Q$ and $R$ acting on $A$. A stronger entropic
uncertainty relation was conjectured by Renes and Boileau \cite{eEUC0}, and
later proved by Berta \textit{et al}. \cite{eEUC}
\begin{equation}
S\left( Q|B\right) +S\left( R|B\right) \geq \log _{2}\frac{1}{c}+S\left(
A|B\right) ,
\end{equation}%
where $S\left( A|B\right) =S(\rho _{AB})-S(\rho _{B})$ is the conditional
von Neumann entropy with $S(\rho )=-$tr$(\rho \log _{2}\rho )$ the von
Neumann entropy \cite{book QIC}. $S\left( X|B\right) $ with $X\in (Q,R)$ is
the conditional von Neumann entropy of the post-measurement state $\rho
_{XB}=\sum_{x}(|\psi _{x}\rangle \left\langle \psi _{x}\right\vert \otimes
\mathbbm{1}%
)\rho _{AB}(|\psi _{x}\rangle \left\langle \psi _{x}\right\vert \otimes
\mathbbm{1}%
)$ after quantum system $A$ is measured by $X$, where \{$|\psi _{x}\rangle $%
\} are the eigenstates of the observable $X$ and $%
\mathbbm{1}%
$ is the identity operator. Although the proof of this
quantum-memory-assisted entropic uncertainty relation is rather complex, the
meaning is clear: the entanglement of systems $A$ and $B$ may lead to a
negative conditional entropy $S(A|B)$ \cite{fushang}, which will in turn
beat the lower bound $\log _{2}\frac{1}{c}$. Especially when $A$ and $B$ are
maximally entangled, the simultaneous measurement of $Q$ and $R$ can be
precisely predicted \cite{eEUC,eEUCs}. In recent, two parallel experiments
\cite{exp1,exp2} have confirmed the quantum-memory-assisted entropic
uncertainty relation.

Quantum objects are inevitably in contact with environments and a
consequence of the interaction is decoherence or dissipation \cite{book
QIC,book open}. So several questions naturally arise: How environmental
noises influence the quantum-memory-assisted entropic uncertainty principle?
Will the noisy channels surely and only increase the amount of uncertainty
because of disentanglement? Is quantum correlation the only key factor for
this uncertainty principle under noise? To answer these questions, we
consider in this Letter two categories of noises: unital and nonunital noisy
channels. Intuition tells us that the uncertainty will increase due to the
noise-induced disentanglement. We demonstrate that it is true for local
unital noises, but may fail for a local amplitude-damping noise, a typical
nonunital channel, in the long-time limit. The mechanism behind this
counterintuitive phenomenon is explored. As a by-product, we present quantum
correlation witness under specific noises by the new uncertainty relation.

We focus on the uncertainty game model illustrated in Ref. \cite{eEUC}: Bob
sends qubit $A$, initially entangled with another qubit $B$ (quantum
memory), to Alice. Then, Alice measures either $Q$ or $R,$ and announces her
measurement choice to Bob. Eq. (1) captures Bob's uncertainty about Alice's
measurement outcome. We assume the two-qubit system to be initially prepared
in a class of states with the maximally mixed subsystems [$\rho _{A(B)}=%
\mathbbm{1}%
^{A(B)}/2$] \cite{Belltai}
\begin{equation}
\rho _{AB}=\frac{1}{4}\left(
\mathbbm{1}%
^{A}\otimes
\mathbbm{1}%
^{B}+\sum_{j=1}^{3}C_{\sigma _{j}}\sigma _{j}^{A}\otimes \sigma
_{j}^{B}\right) ,
\end{equation}%
where $\sigma _{j}$ with $j\in \{1,2,3\}$ are the standard Pauli matrices,
and the coefficients $C_{\sigma _{j}}=$tr$_{AB}(\rho _{AB}\sigma
_{j}^{A}\otimes \sigma _{j}^{B})$ satisfy $0\leq \left\vert C_{\sigma
_{j}}\right\vert \leq 1$. The state of this type is also called
Bell-diagonal state, because it can be diagonalized as a convex combination
of four Bell states: $\rho _{AB}=\lambda _{\Phi ^{+}}|\Phi ^{+}\rangle
\left\langle \Phi ^{+}\right\vert +\lambda _{\Phi ^{-}}|\Phi ^{-}\rangle
\left\langle \Phi ^{-}\right\vert +\lambda _{\Psi ^{+}}|\Psi ^{+}\rangle
\left\langle \Psi ^{+}\right\vert +\lambda _{\Psi ^{-}}|\Psi ^{-}\rangle
\left\langle \Psi ^{-}\right\vert ,$ with eigenstates $|\Phi ^{\pm }\rangle
=\left( |00\rangle \pm |11\rangle \right) /\sqrt{2}$ and $|\Psi ^{\pm
}\rangle =\left( |01\rangle \pm |10\rangle \right) /\sqrt{2},$ and
corresponding eigenvalues $\lambda _{\Phi ^{\pm }}=\left( 1\pm C_{\sigma
_{1}}\mp C_{\sigma _{2}}+C_{\sigma _{3}}\right) /4$ and $\lambda _{\Psi
^{\pm }}=\left( 1\pm C_{\sigma _{1}}\pm C_{\sigma _{2}}-C_{\sigma
_{3}}\right) /4$, respectively. Considering the positivity requirement $%
\lambda _{\Phi ^{\pm }},\lambda _{\Psi ^{\pm }}\geq 0,$ all Bell-diagonal
states should be confined geometrically within a tetrahedron in a
three-dimensional space spanned by $(C_{\sigma _{1}},C_{\sigma
_{2}},C_{\sigma _{3}})$ \cite{Belltai} (See Fig. 1), providing an intuitive
geometric picture for exploring the quantum-memory-assisted entropic
uncertainty principle.

Before investigating the noise effect, we first consider how to consistently
reach the lower bound of Eq. (1). We employ the set of Pauli observables $%
\{\sigma _{j}\}$ with $j\in \{1,2,3\}$. The conditional von Neumann entropy
after qubit $A$ was measured by one of the Pauli observables can be
expressed as $S\left( \sigma _{j}|B\right) =H_{bin}\left( \frac{1+C_{\sigma
_{j}}}{2}\right) ,$ where $H_{bin}\left( p\right) =-p\log _{2}p-(1-p)\log
_{2}(1-p)$ is the binary entropy \cite{book QIC}. Therefore, if we choose
two of the Pauli observables $Q=\sigma _{j}$ and $R=\sigma _{k}$ $(j\neq k)$
for measurement, the left-hand side of Eq. (1) can be written as
\begin{equation}
U=H_{bin}\left( \frac{1+C_{\sigma _{j}}}{2}\right) +H_{bin}\left( \frac{%
1+C_{\sigma _{k}}}{2}\right) .
\end{equation}%
On the other hand, the complementarity $c$ of the observables $\sigma _{j}$
and $\sigma _{k}$ is always equal to 1/2 and the reduced density matrix of
the Bell-diagonal state is a maximally mixed state, i.e., $S(\rho _{B})=1$.
Therefore, the right-hand side of Eq. (1) reduces to $S(\rho _{AB})$ and
takes the form
\begin{equation}
U_{b}=-\sum_{\chi =\Phi ,\Psi ;\varepsilon =\pm }\lambda _{\chi
^{\varepsilon }}\log _{2}\lambda _{\chi ^{\varepsilon }}.
\end{equation}%
In general, Eq. (4) provides a lower bound of uncertainty of Eq. (3). Having
in mind the choice of our measurements $\sigma _{j}$ and $\sigma _{k},$ we
may check that if the initial Bell-diagonal state meets the following
condition
\begin{equation}
C_{\sigma _{i}}=-C_{\sigma _{j}}\cdot C_{\sigma _{k}},\text{ }(i\neq j\neq
k),
\end{equation}%
$U\equiv U_{b}$ will be strictly satisfied in Eq. (1), implying a direct
measurement of the degree of uncertainty by the joint entropy $S(\rho _{AB})$
of the whole system. In what follows, we name Eq. (5) as the state
preparation and measurement choice (SPMC) condition for this new entropic
uncertainty principle.

\begin{figure}[tbp]
\centering
\includegraphics[width=2.4in]{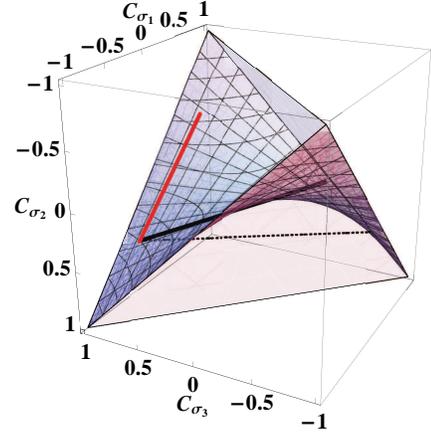}
\caption{(Color online) The geometry of Bell-diagonal states with the blue
(gray) tetrahedron representing the set of all Bell-diagonal states, where
the meshed surface means the valid Bell diagonal states meeting the SPMC
condition $C_{\protect\sigma _{2}}=-C_{\protect\sigma _{1}}\cdot C_{\protect%
\sigma _{3}}$. The black solid, red (gray) solid, and black dotted lines
represent trajectories of Bell-diagonal states ($C_{\protect\sigma _{1}},C_{%
\protect\sigma _{2}},C_{\protect\sigma _{3}}$)=(-0.5,0.4,0.8) under local
bit-flip, phase-flip, and bit-phase-flip noises, respectively. }
\end{figure}

We assume that qubit $A$ will experience a local noisy channel when sent to
Alice, but qubit $B$ is a quantum memory free from noise \cite{quantum
memory}. The evolved state of the whole system can be characterized by the
quantum map $\mathcal{M}\left( \rho _{AB}\right) =\sum_{\mu }\left( \kappa
_{\mu }\otimes
\mathbbm{1}%
\right) \rho _{AB}\left( \kappa _{\mu }\otimes
\mathbbm{1}%
\right) ^{\dag }$ with \{$\kappa _{\mu }$\} the local Kraus operators
satisfying $\sum_{\mu }\kappa _{\mu }^{\dag }\kappa _{\mu }=%
\mathbbm{1}%
.$

We first consider several paradigmatic types of local unital noisy channels:
bit-flip, bit-phase-flip, and phase-flip (equivalent to phase damping),
satisfying the unital condition: $\Lambda _{u}^{A}\left( \frac{1}{2}%
\mathbbm{1}%
^{A}\right) =\frac{1}{2}%
\mathbbm{1}%
^{A}$ \cite{channel}$,$ with $\Lambda _{u}^{A}\left( \rho _{A}\right)
=\sum_{\mu }\kappa _{\mu }\rho _{A}\kappa _{\mu }^{\dag }$. The
corresponding Kraus operators are denoted by $\kappa _{0}^{l}=\sqrt{1-\eta
_{l}}%
\mathbbm{1}%
$, $\kappa _{1}^{l}=\sqrt{\eta _{l}}\sigma _{l}$ with $l=1,2,3$ representing
bit-flip, bit-phase-flip, and phase-flip channels respectively (in the
following, bit-flip, bit-phase-flip, and phase-flip noises are also called $%
\Sigma _{1},\Sigma _{2},\Sigma _{3}$ noises, respectively), and $\eta _{l}$
represents the probability of the noise taking place. It is convenient to
check that the state of qubits $A$ and $B$ initially prepared in a
Bell-diagonal state $(C_{\sigma _{1}},C_{\sigma _{2}},C_{\sigma _{3}})$ will
still be of a Bell-diagonal type when passing through one of the three noisy
channels: $\mathcal{M}\left( \rho _{AB}\right) =\sum_{\chi =\Phi ,\Psi
;\varepsilon =\pm }\lambda _{\chi ^{\varepsilon }}^{^{\prime }}|\chi
^{\varepsilon }\rangle \left\langle \chi ^{\varepsilon }\right\vert ,$ where
$\lambda _{\Phi ^{\pm }}^{^{\prime }}=[1\pm C_{\sigma _{1}}^{\prime }\mp
C_{\sigma _{2}}^{\prime }+C_{\sigma _{3}}^{\prime }]/4$ and $\lambda _{\Psi
^{\pm }}^{^{\prime }}=[1\pm C_{\sigma _{1}}^{\prime }\pm C_{\sigma
_{2}}^{\prime }-C_{\sigma _{3}}^{\prime }]/4$ \cite{Bellstate2} with the
three parameters
\begin{eqnarray}
C_{\sigma _{l}}^{\prime } &=&C_{\sigma _{l}},\text{ }  \notag \\
C_{\sigma _{m}}^{\prime } &=&(1-2\eta _{l})C_{\sigma _{m}},\text{ }(m\neq l).
\end{eqnarray}%
Here $l=1,2,3$ represent qubit $A$ suffering from $\Sigma _{1},\Sigma
_{2},\Sigma _{3}$ noisy channels, respectively. Assuming qubit $A$ will
experience one of the above noises, one can find that for quantum states
initially prepared in Bell-diagonal states meeting the SPMC condition $%
C_{\sigma _{i}}=-C_{\sigma _{j}}\cdot C_{\sigma _{k}}$, the
quantum-memory-assisted entropic uncertainty of the observables $\sigma _{j}$
and $\sigma _{k}$ can consistently reach the lower bound, if no $\Sigma _{i}$
noise takes place \cite{Note}.

\begin{figure}[tbp]
\centering
\includegraphics[width=3in]{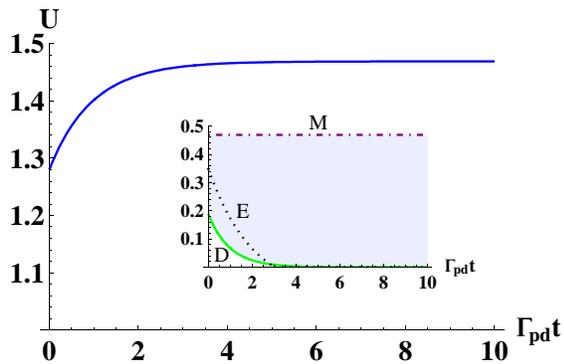}
\caption{(Color online) Uncertainty of the observables $\protect\sigma _{1}$
and $\protect\sigma _{3}$ with initial state ($C_{\protect\sigma _{1}},C_{%
\protect\sigma _{2}},C_{\protect\sigma _{3}}$)=(-0.5,0.4,0.8) under the
local phase-damping channel with $\Gamma _{pd}$ the damping rate. The inset:
entanglement (E), discord (D), and minimal missing information about $A$
after $B$ is measured (M) versus $\Gamma _{pd}t$.}
\label{fig2}
\end{figure}
\textit{\ }\ \ \ \ \ \ \ \ \ \ \ \ \ \ \ \ \ \ \ \ \ \ \ \ \ \ \ \ \ \ \ \ \

For an illustration, the geometric picture of Bell-diagonal states
satisfying the SPMC condition $C_{\sigma _{2}}=-C_{\sigma _{1}}\cdot
C_{\sigma _{3}}$ is depicted as the meshed surface in Fig. 1. On this
surface, the uncertainty of the observables $\sigma _{1}$ and $\sigma _{3}$
is equal to $S(\rho _{AB})$. The black solid, red (gray) solid, and black
dotted lines represent, respectively, trajectories of the Bell-diagonal
state, initially prepared in $\left( C_{\sigma _{1}},C_{\sigma
_{2}},C_{\sigma _{3}}\right) =(-0.5,0.4,0.8)$, under $\Sigma _{1}$, $\Sigma
_{3}$, and $\Sigma _{2}$ noises. Apparently, $\Sigma _{1}$ and $\Sigma _{3}$
noises will not break the SPMC condition, since their trajectories are
always on the surface, whereas $\Sigma _{2}$ noise, due to the departure
from the surface, will definitely break the SPMC condition (except for two
points). As a result, the SPMC condition will be consistently satisfied if
no $\Sigma _{2}$ noise takes place.

To explore the influence of quantum correlations on this uncertainty
relation, we relate the lower bound of Eq. (1) to the definition of discord
(a measure for quantum correlations): $D=-S(A|B)+\min_{\{B_{k}\}}%
\sum_{k}q_{k}S(\rho _{A}^{k})$ \cite{discord}$,$ where min$%
_{\{B_{k}\}}\sum_{k}q_{k}S(\rho _{A}^{k})$ (denoted by $M$ in the following)
captures the minimal missing information about $A$ after $B$ is measured,
and $\rho _{A}^{k}=$tr$_{B}\{B_{k}\rho _{AB}B_{k}^{\dag }\}/q_{k}$ is the
resulting state after the complete measurement \{B$_{k}$\} on qubit $B$, and
$q_{k}=$tr$_{AB}\{B_{k}\rho _{AB}B_{k}^{\dag }\}.$ Therefore, we have
\begin{equation}
U\geq \log _{2}\frac{1}{c}+M-D.
\end{equation}%
For Bell-diagonal states, $M$ can be expressed as \cite{Luo}
\begin{equation}
M=H_{bin}\left( \frac{1+C_{\max }}{2}\right) ,
\end{equation}%
with $C_{\max }=$max$\{|C_{\sigma _{1}}^{\prime }|,|C_{\sigma _{2}}^{\prime
}|,|C_{\sigma _{3}}^{\prime }|\}.$ According to Eq. (6), if $\Sigma _{i}$
noise takes place, as long as $|C_{\sigma _{i}}|\geq |C_{\sigma
_{j}}|,|C_{\sigma _{k}}|,$ we may have $M=H_{bin}\left( \frac{1+|C_{\sigma
_{i}}|}{2}\right) ,$ which is a constant, and implies that the uncertainty
is fully dependent on the quantum correlations between qubit $A$ and quantum
memory $B.$ Especially, if initial state is prepared according to SPMC
condition $C_{\sigma _{j}}=-C_{\sigma _{i}}\cdot C_{\sigma _{k}}$ (or $%
C_{\sigma _{k}}=-C_{\sigma _{i}}\cdot C_{\sigma _{j}}),$ the equality in Eq.
(7) can be consistently satisfied, which suggests that measuring the
uncertainty of the observables $\sigma _{i}$ and $\sigma _{k}$ (or $\sigma
_{i}$ and $\sigma _{j}$) can be directly employed to witness quantum
correlations
\begin{equation}
D=const-U,
\end{equation}%
with $const=\log _{2}\frac{1}{c}+H_{bin}\left( \frac{1+|C_{\sigma _{i}}|}{2}%
\right) $ a constant.

As an example, we consider the phase-damping channel with the Kraus
operators $\kappa _{0}^{pd}=\left\vert 0\right\rangle \left\langle
0\right\vert +e^{-\frac{\Gamma _{pd}t}{2}}\left\vert 1\right\rangle
\left\langle 1\right\vert $ and $\kappa _{1}^{pd}=\sqrt{1-e^{-\Gamma _{pd}t}}%
\left\vert 1\right\rangle \left\langle 1\right\vert ,$ which is equivalent
to the phase-flip channel with $\eta _{3}=\left( 1-e^{-\frac{\Gamma _{pd}t}{2%
}}\right) /2$ \cite{book QIC}$.$ Qubit $A$ and quantum memory $B$ are
initially prepared in a Bell-diagonal state $\left( C_{\sigma
_{1}},C_{\sigma _{2}},C_{\sigma _{3}}\right) =(-0.5,0.4,0.8)$ satisfying the
SPMC condition $C_{\sigma _{2}}=-C_{\sigma _{1}}\cdot C_{\sigma _{3}}$, then
qubit $A$ sent through the phase-damping channel will not break the SPMC
condition, and the uncertainty of the observables $\sigma _{1}$ and $\sigma
_{3}$ can be directly used to detect quantum correlations $D=const-U,$ with $%
const=1+H_{bin}(0.9)$ and $U=H_{bin}(0.9)+H_{bin}(0.5-0.25e^{-\Gamma
_{pd}t/2}).$ As shown in Fig. 2, the uncertainty will increase in the
long-time limit due to the gradually missing quantum correlations.

The dynamics of discord and entanglement (measured by concurrence \cite%
{concurrence}) are shown in the inset of Fig. 2. As we know, entanglement
will not increase under local noisy channels \cite{entanglement}. In
addition, since $\mathcal{M}\left( \rho _{AB}\right) =\sum_{\chi =\Phi ,\Psi
;\varepsilon =\pm }\lambda _{\chi ^{\varepsilon }}^{^{\prime }}|\chi
^{\varepsilon }\rangle \left\langle \chi ^{\varepsilon }\right\vert $ holds
for bit-flip, bit-phase-flip, and phase-flip (phase-damping) noises, we may
conclude that these unital channels are also semiclassical according to the
definition presented in Ref. \cite{channel}. Therefore, quantum correlations
never increase under above local unital channels \cite{channel}. However,
are we sure that the decrease of quantum correlations including entanglement
and discord make the outcomes of two incompatible observables\ more
uncertain?

To further explore this problem, in what follows, we consider a nonunital
and nonsemiclassical local channel, i.e., the amplitude-damping noise with
Kraus operators $\kappa _{0}^{ad}=e^{-\frac{\Gamma _{ad}t}{2}}\left\vert
0\right\rangle \left\langle 0\right\vert +\left\vert 1\right\rangle
\left\langle 1\right\vert $, $\kappa _{1}^{ad}=\sqrt{1-e^{-\Gamma _{ad}t}}%
\left\vert 1\right\rangle \left\langle 0\right\vert $ \cite{book QIC,AD}$.$
Here $\Lambda _{nu}^{A}\left( \frac{1}{2}%
\mathbbm{1}%
^{A}\right) =[e^{-\Gamma _{ad}t}\left\vert 0\right\rangle \left\langle
0\right\vert +(2-e^{-\Gamma _{ad}t})\left\vert 1\right\rangle \left\langle
1\right\vert ]/2$ is not maximally mixed, which implies that the state
through a noisy channel to be measured by Alice is not a Bell-diagonal
state, and the SPMC condition presented above is not satisfied anymore. We
may only study the lower bound of uncertainty instead. Provided the initial
condition $|C_{\sigma _{1}}|\geq |C_{\sigma _{2}}|,$ $M$ can be expressed as%
\begin{equation}
M=\min \{M_{x},M_{z}\},
\end{equation}%
where $M_{x}=H_{bin}(\frac{1+u}{2})$ with $u=\sqrt{e^{-\Gamma
_{ad}t}[C_{\sigma _{1}}^{2}+2\cosh (\Gamma _{ad}t)-2]}$ and $M_{z}=\frac{%
H_{bin}(v_{+})+H_{bin}(v_{-})}{2}$ with $v_{\pm }=(1\pm C_{\sigma _{3}})\exp
(-\Gamma _{ad}t)/2$ \cite{Supplement}, which is time-dependent and may also
be non-monotonic. Therefore, in contrast to Eq. (9), the uncertainty
relation here cannot be directly employed to detect quantum correlations.
Nevertheless, Fig. 3 demonstrates a counterintuitive phenomenon: the
uncertainty of two incompatible observables might be reduced under the
influence of the amplitude-damping noise in the long-time limit (for a
general discussion see in \cite{Supplement}). The key factor for the
uncertainty reduction should not be the quantum correlations, which are
decreasing in this case. Therefore, the quantum correlation is not the only
decisive factor for the amount of the uncertainty.

\begin{figure}[tbp]
\centering
\includegraphics[width=3in]{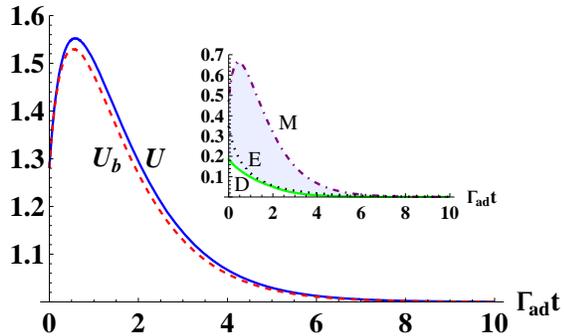}
\caption{(Color online) $U$ and $U_{b}$ of the observables $\protect\sigma %
_{1}$ and $\protect\sigma _{3}$ with initial state ($C_{\protect\sigma %
_{1}},C_{\protect\sigma _{2}},C_{\protect\sigma _{3}}$)=(-0.5,0.4,0.8) under
the local amplitude-damping channel with $\Gamma _{ad}$ the damping rate.
The inset: same parameters as in Fig. 2.}
\label{fig3}
\end{figure}

Now we summarize the above phenomena: (i) The uncertainty (or lower bound)
will increase under local unital noisy channels while it may be reduced
under the nonunital noise channel. (ii) The relation between quantum
correlations and uncertainty is subtle, since the reduced uncertainty
happens in the case that quantum correlations, including discord and
entanglement are also reduced under above local amplitude-damping channel.

The first phenomenon can be explained mathematically. We assume the
bipartite system is initially prepared with the maximally mixed subsystems.
Therefore, in general, we have $U_{b}=\log _{2}\frac{1}{c}+S\left(
A|B\right) =\log _{2}\frac{1}{c}+S\left( \rho _{AB}\right) -\log _{2}d_{B}$ (%
$d_{B}$ the dimension of quantum memory $B$)$.$ As the joint entropy will
not decrease under local unital channels: $S[\mathcal{M}_{lu}(\rho
_{AB})]\geq S(\rho _{AB})$, $U_{b}$ will increase under local unital noise
\cite{Supplement}, however, it may not be true for local nonunital channels.

In order to understand the physical origin of above phenomena, especially
the second one, we reconsider the treatment in Eq. (7). Apparently, the
uncertainty is related to the discrepancy between $M$ and $D$, not just the
quantum correlations only, and it is decided by the competition between
quantum correlations and the minimal missing information of single particle
after local measurement on another one. For illustration, the competition
between $M$ and $D$ are depicted with blue (gray) shades in the insets of
Figs. 2 and 3. Quantum correlations, including entanglement and discord,
will decrease in both cases. However, the most difference is that Eq. (10)
is not a monotonic function, which may even decrease under amplitude-damping
channel. That is to say, the missing information by local measurements may
be reduced in the long-time limit, which in turn lowers the uncertainty.

The work presented above can be immediately investigated by all-optical
setups, where the state preparation and measurement can be realized like in
Refs. \cite{exp1,exp2} and the noisy channels can be simulated according to
Refs. \cite{exp3,exp4}\emph{.}

Before ending this Letter, we mention some open problems waiting for
solution: What is the physical mechanism behind the competition between M
and D in a general frame? Will this counterintuitive phenomenon exist in all
nonunital noisy channels and for all quantum states? Is it possible to find
the minimum uncertainty achievable in the presence of nonunital noise? Can
we find similar phenomena by other entropy measures, e.g., the uncertainty
relation based on smooth entropy \cite{smooth}? Most fascinatingly, is it
possible to directly utilize the decoherence or dissipation properties
illustrated in this Letter to perform quantum information tasks such as
quantum key distribution?

In summary, we have studied the noise effect on quantum-memory-assisted
entropic uncertainty principle. By investigating different noises, we have
demonstrated that local unital noises will surely increase the uncertainty,
but under the influence of a nonunital amplitude-damping channel, we found
remarkably that the uncertainty might even be reduced. Our work is the first
step toward the study of the noise effect on the quantum-memory-assisted
entropic uncertainty principle and can be practically estimated by quantum
tomography with currently experimental setups in \cite{exp1,exp2,exp3,exp4}.

\vspace{0.1cm} This work is supported by the National Natural
Science Foundation of China under Grant No. 10974225 and No.
11004226, and by the National Fundamental Research Program of China
under Grant No. 2012CB922102. Z.Y.X. also acknowledges support by
Soochow University Startup Fund under Grant No. Q410800911.

Note added: After finishing this work, we became aware of several
topics on the role of quantum correlations in the entropic
uncertainty principle, such as arXiv:1202.0939 and arXiv:1203.3153.

\vspace{0cm}

\begin{center}
\newpage \textbf{Supplemental Material}
\end{center}

\subsection{I. The SPMC condition of Eq. (5)}

For clarity, we denote Eq. (4) by $U_{b}=-\sum_{x,y=0,1}\lambda _{xy}\log
_{2}\lambda _{xy}$, with $\lambda _{xy}=[1+(-1)^{x}C_{\sigma
_{1}}-(-1)^{x+y}C_{\sigma _{2}}+(-1)^{y}C_{\sigma _{3}}]/4$ the eigenvalues
of a Bell-diagonal state. We first consider one of the SPMC conditions,
e.g., $C_{\sigma _{1}}=-C_{\sigma _{2}}\cdot C_{\sigma _{3}}.$ Then $\lambda
_{xy}=\left[ 1-(-1)^{x+y}C_{\sigma _{2}}\right] \cdot \lbrack
1+(-1)^{y}C_{\sigma _{3}}]/4$ and
\begin{eqnarray*}
U_{b} &=&-\sum_{x,y=0,1}\{\frac{[1-(-1)^{x+y}C_{\sigma _{2}}]\cdot \lbrack
1+(-1)^{y}C_{\sigma _{3}}]}{4} \\
&&\log _{2}\frac{1-(-1)^{x+y}C_{\sigma _{2}}}{2}\} \\
&&-\sum_{x,y=0,1}\{\frac{[1-(-1)^{x+y}C_{\sigma _{2}}]\cdot \lbrack
1+(-1)^{y}C_{\sigma _{3}}]}{4} \\
&&\log _{2}\frac{1+(-1)^{y}C_{\sigma _{3}}}{2}\} \\
&=&-\frac{1-C_{\sigma _{2}}}{2}\log _{2}\frac{1-C_{\sigma _{2}}}{2}-\frac{%
1+C_{\sigma _{2}}}{2}\log _{2}\frac{1+C_{\sigma _{2}}}{2}%
~~~~~~~~~~~~~~~~~~~~~~~~ \\
&&-\frac{1-C_{\sigma _{3}}}{2}\log _{2}\frac{1-C_{\sigma _{3}}}{2}-\frac{%
1+C_{\sigma _{3}}}{2}\log _{2}\frac{1+C_{\sigma _{3}}}{2}%
~~~~~~~~~~~~~~~~~~~~~~~~ \\
&=&H_{bin}\left( \frac{1+C_{\sigma _{2}}}{2}\right) +H_{bin}\left( \frac{%
1+C_{\sigma _{3}}}{2}\right) ,~~~~~~~~~~~~~~~~~~~~~~~~~~~~~~~~~~~~
\end{eqnarray*}%
which is equal to the uncertainty of the observables $\sigma _{2}$ and $%
\sigma _{3}$ [Eq. (3)]$.$ For other two SPMC conditions, the proof is
similar.

\subsection{II. Expression of Eq. (10)}

Although a Bell-diagonal state under local amplitude-damping channel will no
longer be of the Bell-diagonal type, it is still of \textquotedblleft
X\textquotedblright\ type%
\begin{equation*}
\rho _{AB}(t)=\frac{1}{2}\left(
\begin{array}{llll}
v_{+} & 0 & 0 & w_{-} \\
0 & v_{-} & w_{+} & 0 \\
0 & w_{+} & 1-v_{+} & 0 \\
w_{-} & 0 & 0 & 1-v_{-}%
\end{array}%
\right) ,
\end{equation*}%
with $v_{\pm }=e^{-\Gamma _{ad}t}(1\pm C_{\sigma _{3}})/2$ and $w_{\pm
}=e^{-\Gamma _{ad}t/2}(C_{\sigma _{1}}\pm C_{\sigma _{2}})/2.$ In general,
we may employ projectors $\{B_{m}\}=\{\cos \theta \left\vert 0\right\rangle
+e^{i\xi }\sin \theta \left\vert 1\right\rangle$, $e^{-i\xi }\sin \theta
\left\vert 0\right\rangle -\cos \theta \left\vert 1\right\rangle \}$ \cite%
{smII-1}. If $|w_{+}+w_{-}|\geq |w_{+}-w_{-}|$, i.e., $|C_{\sigma _{1}}|\geq
|C_{\sigma _{2}}|$, the optimal measurement is either \{($\left\vert
0\right\rangle +\left\vert 1\right\rangle )/\sqrt{2},(\left\vert
0\right\rangle -\left\vert 1\right\rangle )/\sqrt{2}$\}, i.e., $\sigma _{x}$
operation or \{$\left\vert 0\right\rangle ,-\left\vert 1\right\rangle $\},
i.e., $\sigma _{z}$ operation \cite{smII-2}. Therefore, the minimal missing
information after local measurement can be expressed as $M=\min
\{M_{x},M_{z}\},$ where $M_{x}=H_{bin}(\frac{1+u}{2})$ with $u=\sqrt{%
e^{-\Gamma _{ad}t}[C_{\sigma _{1}}^{2}+2\cosh (\Gamma _{ad}t)-2]}$ and $%
M_{z}=\frac{H_{bin}(v_{+})+H_{bin}(v_{-})}{2}$. With $M$, the discord can be
calculated straightforwardly.

\begin{figure*}[tbp]
\centering
\includegraphics[width=4in]{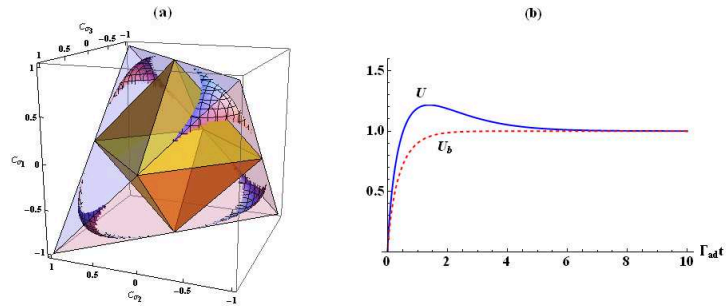}
\renewcommand{\figurename}{Figure.SM} \caption{(Color online) (a) The geometry of Bell-diagonal
states. The yellow octahedron is the set of separable states while
the four little blue tetrahedrons are set of entangled states. The
four meshed curve surfaces correspond to the critical condition for
the decrease (inside) or increase (outside) of uncertainty (lower
bound) under local nonunitial amplitude-damping channel in the
long-time limit. (b) A trivial example: The
initial state prepared in ($C_{\protect\sigma _{1}},C_{\protect\sigma %
_{2}},C_{\protect\sigma _{3}}$)=(-1,1,1) will increase the
uncertantiy (lower bound) in the long-time limit.}
\end{figure*}

\subsection{III. The role of the initial state for the decrease of
uncertainty under nonunital amplitude-damping channel}

Will all initial states prepared in Bell-diagonal type under the influence
of local nonunital amplitude-damping noise decrease the uncertainty (lower
bound) in the long-time limit? The answer is no. Actually, only if we
prepare the initial Bell-diagonal state satisfying the condition $%
U_{b}(0)>U_{b}(\infty ),$ will the phenomenon illustrated in the
manuscript arise. To see more clearly, a geometry for the condition
is depicted in Figure.SM(a): initial states inside the four meshed
curve surfaces are those satisfying the condition. As a trivial
example (for the uncertainty lower
bound increasing under the noise), we consider one of the four Bell states $%
\left( C_{\sigma _{1}},C_{\sigma _{2}},C_{\sigma _{3}}\right)
=(-1,1,1)$, i.e., the vertex of the tetrahedron, as the initial
state. Figure.SM(b) shows that the uncertainty (lower bound) will
increase in the long-time limit compared with the situation when no
noise takes place.

\subsection{IV. The proof of increasing uncertainty under local unital noisy
channels}

\textit{Proof}. Consider a bipartite system AB with dimension $%
d_{AB}=d_{A}\times d_{B}.$ We first demonstrate that if the local noisy
channel $\Lambda _{u}^{A}$ is unital, then the map $\mathcal{M}_{lu}=\left[
\Lambda _{u}^{A}\otimes
\mathbbm{1}%
^{B}\right] $ is still unital. The proof is straightforward: $\mathcal{M}%
_{lu}\left( \frac{%
\mathbbm{1}%
^{AB}}{d_{AB}}\right) =\sum_{\mu }\left( \kappa _{\mu }\otimes
\mathbbm{1}%
\right) \frac{%
\mathbbm{1}%
^{AB}}{d_{AB}}\left( \kappa _{\mu }\otimes
\mathbbm{1}%
\right) ^{\dag }$ $=\sum_{\mu }\left( \kappa _{\mu }\frac{%
\mathbbm{1}%
^{A}}{d_{A}}\kappa _{\mu }^{\dag }\right) \otimes \frac{%
\mathbbm{1}%
^{B}}{d_{B}}=\frac{%
\mathbbm{1}%
^{A}}{d_{A}}\otimes \frac{%
\mathbbm{1}%
^{B}}{d_{B}}=\frac{%
\mathbbm{1}%
^{AB}}{d_{AB}}.$

Recalling $U_{b}=\log _{2}\frac{1}{c}+S\left( A|B\right) =\log _{2}\frac{1}{c%
}+S\left( \rho _{AB}\right) -S(\rho _{B}).$ If the state is initially
prepared in the maximally mixed subsystems, i.e., $\rho _{A}=\frac{%
\mathbbm{1}%
^{A}}{d_{A}}$ $(\rho _{B}=\frac{%
\mathbbm{1}%
^{B}}{d_{B}}),$ then quantum memory $B$ free from noise will keep maximally
mixed: $S(\rho _{B})\equiv \log _{2}d_{B}$ \cite{smIII-1}$.$ Therefore $%
U_{b} $ is fully dependent on $S\left( \rho _{AB}\right) $ and our task now
is to prove $S\left[ \mathcal{M}_{lu}(\rho _{AB})\right] \geq S\left( \rho
_{AB}\right) .$ The proof can be with the help of the monotonicity of the
relative entropy for quantum maps: $S\left[ \mathcal{M}(\rho _{AB})||%
\mathcal{M}(\frac{%
\mathbbm{1}%
^{AB}}{d_{AB}})\right] \leq S\left[ \rho _{AB}||\frac{%
\mathbbm{1}%
^{AB}}{d_{AB}}\right] $ \cite{smIII-2}$.$ Then we have $-S\left[ \mathcal{M}%
(\rho _{AB})\right] -$tr$\left[ \mathcal{M}(\rho _{AB})\log _{2}\mathcal{M}%
\left( \frac{%
\mathbbm{1}%
^{AB}}{d_{AB}}\right) \right] \leq -S\left( \rho _{AB}\right) -$tr$\left[
\rho _{AB}\log _{2}\frac{%
\mathbbm{1}%
^{AB}}{d_{AB}}\right] =-S\left( \rho _{AB}\right) +\log _{2}d_{AB}.$ If $%
\mathcal{M}$ is unital, then tr$\left[ \mathcal{M}_{lu}(\rho _{AB})\log _{2}%
\mathcal{M}_{lu}\left( \frac{%
\mathbbm{1}%
^{AB}}{d_{AB}}\right) \right] =$tr$\left[ \mathcal{M}_{lu}(\rho _{AB})\log
_{2}\frac{%
\mathbbm{1}%
^{AB}}{d_{AB}}\right] =-\log _{2}d_{AB}.$ Finally, we may obtain $S\left[
\mathcal{M}_{lu}(\rho _{AB})\right] \geq S\left( \rho _{AB}\right),$ which
implies that $U_{b}$ will not decrease under local unital noisy channels.
However, this conclusion does not apply to local nonunital noisy channels. \
\ \ \ \ \ \ $\square $


\begin{thebibliography}{99}
\bibitem{Note} Given $\Sigma _{i}$ noise takes place, we may have $C_{\sigma
_{i}}^{\prime }=C_{\sigma _{i}},$ $C_{\sigma _{j}}^{\prime }=(1-2\eta
_{i})C_{\sigma _{j}},$ and $C_{\sigma _{k}}^{\prime }=(1-2\eta
_{i})C_{\sigma _{k}}$ according to Eq. (6). Assuming that the SPMC condition
can still be consistently satisfied under $\Sigma _{i}$ noise, i.e., $%
C_{\sigma _{i}}^{\prime }=-C_{\sigma _{j}}^{\prime }\cdot C_{\sigma
_{k}}^{\prime }$, we may get the relation $C_{\sigma _{i}}=-C_{\sigma
_{j}}\cdot C_{\sigma _{k}}\cdot $ $(1-2\eta _{i}$)$^{2},$ which requires $%
(1-2\eta _{i}$)$^{2}=1,$ i.e., $\eta _{i}=0$ or $1$ (just two extreme cases).
\end{thebibliography}
\end{document}